\documentclass[11pt]{article}
\usepackage{moriond,epsfig}

\def\be{\begin{equation}}
\def\ee{\end{equation}}
\def\bea{\begin{eqnarray}}
\def\eea{\end{eqnarray}}

\def\Black{}
\def\Blue{}
\def\Brown{}
\def\slash#1{\setbox0=\hbox{$#1$}#1\hskip-\wd0\dimen0=5pt\advance
\dimen0 by-\ht0\advance\dimen0 by\dp0\lower0.5\dimen0\hbox
to\wd0{\hss\sl/\/\hss}}
\begin{document}

\begin{titlepage}
\null
\vspace{3cm}
\begin{center}
\Large\bf 
\Brown
$B \to \rho \pi$ and the unitary angle $\alpha$
\Black
\end{center}
\vspace{1.5cm}

\begin{center}
Aldo Deandrea\\
\vspace{0.5cm}
Theory Division, CERN, CH-1211 Gen\`eve 23, Switzerland
\end{center}

\vspace{1.3cm}

\begin{center}
\Brown
{\bf Abstract}\\[0.5cm] \Black
\parbox{14cm}{The role of the decay mode $B\to\rho\pi$ for the
determination of the unitarity angle $\alpha$ is critically 
examined in view of the smaller than expected ratio
${\cal B}(B \to \rho^\pm\pi^\mp)/{\cal B}(B \to \rho^0\pi^\pm )$
found by the CLEO collaboration. }
\end{center}

\vspace{2cm}

\begin{center}
\Blue
{\sl To appear in the Proceedings of the XXXVth Rencontres de Moriond\\
Electroweak Interactions and Unified Theories\\
Les Arcs 1800, France, March 11-18 2000.}
\Black
\end{center}

\vfil
\noindent
\Brown
CERN-TH/2000-128\\
May 2000
\Black

\end{titlepage}

\newpage
\setcounter{page}{1}

\vspace*{4cm}
\begin{center}
\Large
$B \to \rho \pi$ and the unitary angle $\alpha$
\end{center}
\vspace{0.5cm}

\begin{center}
Aldo Deandrea\\
\vspace{0.5cm}
Theory Division, CERN, CH-1211 Gen\`eve 23, Switzerland
\end{center}

\vspace{0.5cm}

\begin{center}
{\bf Abstract}\\[0.5cm]
\parbox{14cm}{The role of the decay mode $B\to\rho\pi$ for the
determination of the unitarity angle $\alpha$ is critically 
examined in view of the smaller than expected ratio
${\cal B}(B \to \rho^\pm\pi^\mp)/{\cal B}(B \to \rho^0\pi^\pm )$
found by the CLEO collaboration.}
\end{center}

\vspace{2cm}

\section{Introduction}
The task of determining the angle $\alpha$ is complicated by the problem 
of separating two different weak hadronic matrix elements, each carrying 
its own weak phase.
The evaluation of these contributions, referred to in the literature as the 
{\it tree} ($T$) and the {\it penguin} ($P$) contributions, suffers from the
common theoretical uncertainties related to the estimate of composite
four-quark operators between hadronic states. For these estimates, 
only approximate schemes, such as the factorisation approximation, exist 
at the moment, and for this reason several ingenuous schemes have been devised,
trying to disentangle $T$ and $P$ contributions. In general one tries to
exploit the fact that in the $P$ amplitudes only the isospin--$1/2$ 
\footnote{If one neglects electroweak penguins.} part of
the non--leptonic Hamiltonian is active in the decay $B \to \pi \pi$ 
\cite{gronau}; by measurements involving several different isospin amplitudes,
one can separate the two amplitudes and get rid of the
ambiguities arising from the ill--known penguin matrix elements. However
such measurements are not simple, for example $B \to \pi^0 \pi^0$ is probably
small and difficult to detect; the recent measurement by CLEO \cite{cleobpipi}
of the $B \to \pi^+ \pi^-$ branching ratio shows that penguin contributions
are certainly not small; discrete ambiguities reduce the predictive power 
of the isospin method.

In order to have an alternative measurement of the angle $\alpha$ 
different strategies were proposed, either involving 
all the decay modes of a $B$ into a
$\rho\pi$ pair as well as three time--asymmetric quantities
measurable in the three channels for neutral $B$ decays, 
or attempting to measure only the neutral $B$ decay modes by looking 
at the time-dependent asymmetries in different regions of the Dalitz 
plot \cite{lipkin,snyder}.

Preliminary to these analyses is the assumption that, using cuts in the
three invariant masses for the pion pairs, one can extract the $\rho$
contribution without significant background contaminations. The
$\rho$ has spin $1$, the $\pi$ spin $0$ as well as the initial $B$, and
therefore the $\rho$ has angular distribution $\cos ^2 \theta$
($\theta$ is the angle of one of the $\rho$ decay products with the
other $\pi$ in the $\rho$ rest frame). This means that the Dalitz plot
is mainly populated at the border, especially the corners, by this
decay. Only very few events should be lost by excluding the interior of the
Dalitz plot, which is considered a good way to exclude or at least
reduce backgrounds. Analyses following these hypotheses were
performed by the BaBar working groups \cite{babar}; MonteCarlo
simulations, including the background from the $f_0$ resonance, show
that, with cuts at $m_{\pi\pi}=m_\rho\pm 300$ MeV, no significant contributions
from other sources are obtained. Also the role of excited resonances such as
the $\rho^\prime$ and the non--resonant background has been discussed
\cite{charles}.

A signal of possible difficulties for this strategy arises from new results
from the CLEO Collaboration \cite{gao}: 
\begin{equation} {\mathcal B}( B^\pm~\to~\rho^0\pi^\pm)~=~
(10.4^{+3.3}_{-3.4}\pm 2.1)\times 10^{-6}~,\label{eq:12}
\end{equation}
\begin{equation}
{\mathcal B}( B~\to~\rho^\mp\pi^\pm)~=~(27.6^{+8.4}_{-7.4}\pm
4.2)\times 10^{-6}~,\label{eq:13}
\end{equation} 
with a ratio
\begin{equation}\label{eq:88}
R~=~\frac{ {\mathcal B}( B~\to~\rho^\mp\pi^\pm) }{ {\mathcal B}(
B^\pm~\to~\rho^0\pi^\pm)}~=~2.65\pm 1.9~.
\end{equation}
As discussed in \cite{gao}, this ratio looks rather small; as a matter of
fact, when computed in simple approximation schemes, as
factorisation with no penguins, one gets, from the WBS model \cite{WBS}, 
$R \simeq 6$ (using $a_1=1.02$, $a_2=0.14$). The inclusion of penguin 
contributions in the factorisation approximation does not help explaining
the experimental result as one gets $R \simeq 5.5$. One may wonder if
the factorisation approximation is too rough to give an accurate result
or if penguin contributions are larger than expected, in order to explain the
experimental result. An estimate of non--factorisable contributions was
given by Martinelli and co--workers \cite{martinelli} by parameterising
these contributions in terms of the known $B\to \pi K$ decays. Using their 
estimates we deduced values for the ratio $R$ (see Table 1) and they all
agree with the result obtained in the factorisation approximation. Therefore 
some other reason may be advocated for explaining the experimental result
for the ratio $R$.

\begin{table} [h]
\caption{Estimates for the ratio $R$ beyond the factorisation approximation
(the so--called charming penguins) using different sets of input data: 
QCD sum--rules (QCDSR), lattice--QCD, quark models (QM)}
 \vspace{0.4cm}
\begin{center}
\begin{tabular}{|cc|}
\hline
QCDSR & $R=6.3$ \\
lattice & $R=5.5$ \\
QM & $R=6.4$ \\
\hline
\end{tabular}
\end{center}
\end{table}

In \cite{alphanoi} we showed that a new contribution,
not discussed before, is indeed relevant to the decay
of a charged $B$ to $\rho \pi$ and to a lesser extent to the decay of
a neutral $B$ to $\rho \pi$. It arises from the virtual resonant production
depicted in Fig. 1, where the intermediate particle is the $B^*$
meson resonance or other excited states. 
\begin{figure} [h]
\epsfxsize=6cm
\centerline{\epsffile{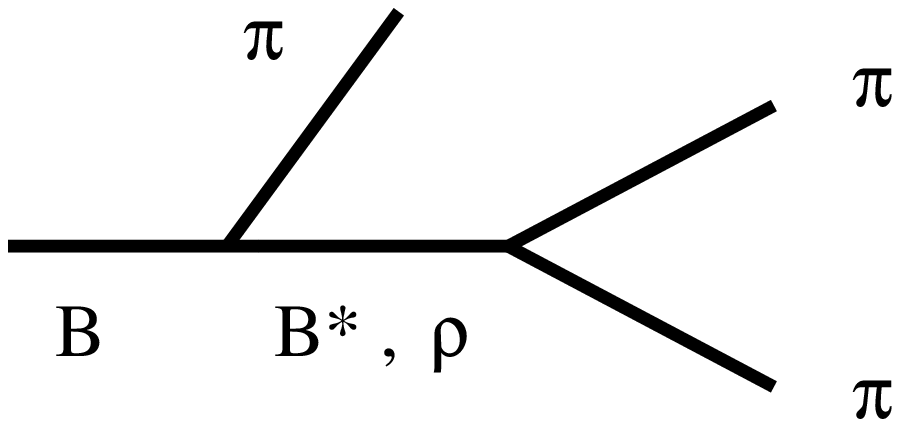}}
\noindent
{\bf Fig. 1} - {The polar diagram. For the $B$ resonances ($B^*=1^-$, $0^+$) 
the strong coupling is on the left and the weak coupling on the right; the 
situation is reversed for the $\rho$ production.}
\end{figure}
The $B^*$ resonance, 
because of phase--space limitations, cannot be produced on the mass shell.
Nonetheless the $B^*$ contribution might be important, owing to its almost
degeneracy in  mass with the $B$ meson; therefore its tail 
may produce sizeable effects in some of the decays of $B$
into light particles, also because it is known theoretically that the
strong coupling constant between $B$, $B^*$ and a pion is
large \cite{gatto}. Concerning other states, we expect their role to
decrease with their mass, since there is no enhancement from the virtual
particle propagator; we shall only consider the $0^+$ state $B_0$ with
$J^P=0^+$ because its coupling to a pion and the meson $B$ is known
theoretically to be uniformly (in momenta) large \cite{gatto}.

\section{Interaction Hamiltonian}
The effective weak non-leptonic Hamiltonian for the $|\Delta
B|=1$ transition is \footnote{We omit, as 
usual in these analyses, the electroweak operators
$Q_k$ ($k=$ 7, 8, 9, 10); they are in general small, but for 
$Q_9$, whose role might be sizeable; its inclusion in
the present calculations would be straightforward.}:
\be
H~=~\frac{G_F}{\sqrt 2}\left\{ V^*_{ub}V_{ud}\sum_{k=1}^2
C_k(\mu)Q_k~-~V^*_{tb}V_{td}\sum_{k=3}^6
C_k(\mu)Q_k\right\}~.\ee
We use the following values of the Wilson coefficients:
$C_1=-0.226$, $C_2=1.100$, $C_3=0.012$, $C_4=-0.029$, $C_5=0.009$,
$C_6=-0.033$; they are obtained in the HV scheme \cite{buras}, with
$\Lambda^{(5)}_{\bar{MS}}=225$ MeV, $\mu={\bar m}_b(m_b)=4.40$ GeV and
$m_t=170 $ GeV.
For the CKM mixing matrix we use the Wolfenstein parameterisation
with $\rho=0.05$, $\eta = 0.36$ and $A=0.806$ in the 
approximation accurate to order $\lambda^3$ in the real part and
$\lambda^5$ in the imaginary part, i.e. $V_{ud}=1-\lambda^2 /2$,
$V_{ub}=A \lambda^3 \;[\rho - i \eta \; (1-\lambda^2 /2)]$, $V_{td}=
A \lambda^3 (1-\rho - i \eta)$ and $V_{tb}=1$.

The diagram of Fig.~1 describes two processes.
For the $B^*$ intermediate state there is an emission of a  
pion by strong interactions, followed by the weak decay of the virtual
$B^*$ into two pions; for the $\rho$ intermediate state there is a
weak decay of $B \to \rho \pi$ followed by the strong decay of 
the $\rho$ resonance. We compute these diagrams as Feynman graphs
of an effective theory within the factorisation approximation, 
using information from the effective Lagrangian
for heavy and light mesons \cite{report} and form factors for the 
couplings to the weak currents.

\section{$B \to \rho \pi$ Decays}
For the charged $B$ decays we obtain the results in  Table 2, 
with $g=0.40$ and $h=-0.54$, which lie in the middle of the
allowed ranges for these parameters. The
branching ratios are obtained with $\tau_B=1.6$ psec and, by integration
over a limited section of the Dalitz plot, defined as $m_\rho-\delta\leq
({\sqrt t}, {\sqrt t^\prime}) \leq m_\rho+\delta$ for $B^-\to\pi^-\pi^-\pi^+$  
and $m_\rho-\delta\leq ({\sqrt s}, {\sqrt s^\prime} ) \leq m_\rho+\delta$ 
for $B^-\to\pi^-\pi^0\pi^0$. For $\delta$ we take $300$ MeV. This amounts 
to require that two of the three pions (those corresponding to the charge 
of the $\rho$) reconstruct the $\rho$ mass within an interval of $2 \delta$. 

\begin{table}[h]
\caption{Effective branching ratios for the charged $B$ decay channels into 
three pions for the choice of the strong coupling constants $g=0.40$ and 
$h=-0.54$. Cuts as indicated in the text.}
\begin{center}   
\begin{tabular}{|cccc|}
\hline
Channels& $\rho$ & $\rho+B^*$ & $\rho+B^*+ B_0$ \\
\hline
$B^-\to\pi^-\pi^0\pi^0~$ &
$1.0\times 10^{-5}$&
$1.0\times 10^{-5}$&
$1.0\times 10^{-5}$ \\
$B^-\to\pi^+\pi^-\pi^-$& 
$0.41\times 10^{-5}$&
$0.58\times 10^{-5}$&
$0.63\times 10^{-5}$\\
\hline
\end{tabular}
\end{center}
\end{table}

We can notice that the inclusion of the new diagrams 
($B$ resonances in Fig. 1) produces practically no effect
for the $B^-\to\pi^-\pi^0\pi^0~$  decay mode, while for 
$B^-\to\pi^+\pi^-\pi^-~$ the effect is significant. 
In Table 2 the overall effect is an increase 
of $50\%$ of the branching ratio as compared to the result
obtained by the $\rho$ resonance alone. It should be observed that the events
arising from the $B$ resonances diagrams represent an irreducible
background. The
contributions from the $B$ resonances populate the whole Dalitz plot and,
obviously, cutting around $t\sim t^\prime\sim m_\rho$ significantly
reduces them. Nevertheless their effect can survive the experimental cuts,
since there can be enough data at the corners, where the contribution 
from the $\rho$ dominates. Integrating on the whole Dalitz plot, with no cuts
and including all contributions, gives
${\mathrm Br} (B^-\to\pi^-\pi^0\pi^0) = 1.5 \times 10^{-5}$ and 
${\mathrm Br} (B^-\to\pi^+\pi^-\pi^-) = 1.4 \times 10^{-5}$
where the values of the coupling constants are as in Table 2.

We now turn to the neutral $B$ decay modes.
The results in Table 3 show basically no
effect for the $\bar{B}^0\to\rho^\pm\pi^\mp$ decay channels and a moderate
effect for the $\rho^0\pi^0$ decay channel. The effect in this channel 
is of the order of 20\% (resp. 50\%) for ${\bar B}^0$ (resp. $B^0$) decay,  
for the choice $g=0.60, \, h=-0.70$ (the one maximising the effect of the
$B^*$ resonances); for smaller
values of the strong coupling constants the effect is reduced.

\begin{table}[h]
\caption{Effective branching ratios for the neutral $ B$ decay
channels into $\rho\pi$ ($g=0.60,\, h=-0.70$). Cuts as indicated in the text.} 
\begin{center}
\begin{tabular}{|cccc|}
\hline
Channels& $\rho$ & $\rho+B^*$ & $\rho+B^*+ B_0$ \\
\hline
${\bar B}^0\to\rho^-\pi^+~$ &
$0.50\times 10^{-5}$&
$0.52\times 10^{-5}$&
$0.49\times 10^{-5}$ \\
${\bar B}^0\to\rho^+\pi^-~$ &
$1.7\times 10^{-5}$&
$1.7\times 10^{-5}$&
$1.7\times 10^{-5}$ \\
${\bar B}^0\to\rho^0\pi^0~$ &
$0.10\times 10^{-5}$&
$0.15\times 10^{-5}$&
$0.12\times 10^{-5}$ \\ \hline
${B}^0\to\rho^+\pi^-~$ &
$0.49\times 10^{-5}$&
$0.51\times 10^{-5}$&
$0.48\times 10^{-5}$ \\
${B}^0\to\rho^-\pi^+~$ &
$1.7\times 10^{-5}$&
$1.7\times 10^{-5}$&
$1.7\times 10^{-5}$ \\
${B}^0\to\rho^0\pi^0~$ &
$0.11\times 10^{-5}$&
$0.17\times 10^{-5}$&
$0.15\times 10^{-5}$ \\
\hline
\end{tabular}
\end{center}
\end{table}

Integration on the whole Dalitz plot, including all contributions, gives 
${\mathrm Br} (\bar{B}^0\to \pi^+\pi^-\pi^0) = 2.6 \times 10^{-5}$
confirming again that most of the branching ratio is due to the
$\rho$-exchange (the first three lines of the $\rho$
column in Table 3 sum up to $2.3 \times 10^{-5}$).
 
\section{Conclusions}
The effect of including $B$ resonance polar diagrams is significant 
for the $B^\mp \to\pi^\mp \pi^\mp \pi^\pm$ decay
and negligible for the other charged $B$ decay mode. This result is of
some help in explaining the recent results from the CLEO Collaboration,
since we obtain $R = 3.5\pm 0.8$, 
to be compared with the experimental result in eq. (\ref{eq:88}). The $\rho$
resonance alone would produce a result up to a factor of 2 higher. 
Therefore we conclude that the polar diagrams examined \cite{alphanoi} 
are certainly relevant in the study of the charged $B$ decay into three pions.
Concerning the neutral $B$ decays which are relevant for the 
determination of the unitarity angle $\alpha$, only the $\rho^0 \pi^0$ decay
channel is partially affected by the extra contributions we considered.

\end{document}